\documentclass[9pt]{book}

\usepackage[dvips]{graphicx,color}
\usepackage{makeidx,universe}

\makeauthorindex
\BookTitle{The proceedings of the Physics of Accreting Compact Binaries}

\CopyRight{\copyright 2010 by Universal Academy Press, Inc.}

\begin{document} %%%%%%%%%%%%%%%

%\tableofcontents
\pagenumbering{arabic}

\chapter{
Quiet Novae with Flat Maximum -- No Optically Thick Winds
}
%%%%%%%%%% <===== TITLE of the contribution
%%%%%%%%%%% The first letter of each word should be capital letter.
% Manuscripts for\\ Symposium Proceedings}

\author{\raggedright \baselineskip=10pt%
{\bf Mariko Kato$^{1}$
}
\\ %%%%%% <== Authors
{\small \it %
(1) Keio University, Hiyoshi, Yokohama 223-8521, Japan
% (2) Department of Physics, University of Texas at El Paso, El Paso, TX 79968, USA 
}}

%**************************
% Please note:
% One \AuthorContents{} is necessary
%    for EACH CONTRIBUTION, for the contents page and
% One \AuthorIndex{} is necessary
%    for EACH AUTHOR, for the index.
%**************************

\AuthorContents{Mariko Kato} %%%%%%% <=== It is the data for CONTENTS. Please enter all author's name that should be initialized.

\AuthorIndex{Kato}{M.} %%%%%%% <=== It is the data for AUTHOR INDEX. Please enter a author's name that should be initialized.
     \baselineskip=10pt
     \parindent=10pt

\section*{Abstract} %%%%%%%%%%%%%%%

I will explain why most novae show a sharp optical peak in the light curve,  
whereas a small number of novae, such as PU Vul, shows a long-lasted 
flat optical maximum. 
Re-examination of occurrence condition of optically thick winds clarifies 
that hydrostatic evolutions, in which optically thick winds 
are suppressed, are realized during nova outbursts on low mass white dwarfs (WDs) 
(less than $\sim 0.7~M_{\odot}$). 
In such a case, a nova outburst evolves very slowly, because 
of no strong mass-ejection, and stays at low temperature stage a 
long time, which results in a long-lasted flat optical peak. 
This explains outburst nature of symbiotic nova PU Vul, that shows  
flat optical maximum lasted as long as eight years with 
no spectral indication of wind mass-loss.
On the other hand, in ordinary nova outbursts, 
strong optically thick winds inevitably occur that carry out 
most of the envelope matter very quickly. Thus the light curve 
decays immediately after the optical peak, which results in a sharp peak 
of optical light curve.

\begin{figure}[t]%%%%%%   Fig 1 %%%  HR diagram  %%%%%%%%%%%

 \begin{tabular}{cc}
  \begin{minipage}{.65\hsize}
   \begin{center}
     \includegraphics[width=.98\textwidth]{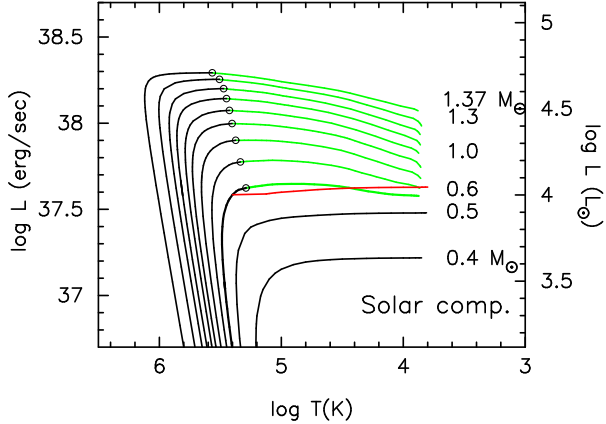}
      \caption{HR diagram for the decay phase of shell flash. 
Optically thick winds occur in the green region and stop at the small 
open circle. No winds occur in $\leq 0.5 M_\odot$ WD.
The WD mass is attached beside the curve. Very massive WDs ($\geq 1.35 _\odot$)
correspond recurrent novae and less massive WDs slow/symbiotic novae.
There are double solutions of wind mass-loss (green) and static (red) in 
the 0.6 $M_\odot$ WD.
 }
    \label{katomariko.f1}
   \end{center}  \end{minipage}  \end{tabular}
\end{figure}%%%%%%%%%%%%%%%

\begin{figure}[t]%%%%%%  Fig 2 and 3 %%%%  RS Oph  %%%%%%%%%%%
 \begin{tabular}{cc}  \begin{minipage}{.40\hsize}   \begin{center}
     \includegraphics[width=.98\textwidth]{f2.epsi}
      \caption{Three ($y$, $V$, and $I_c$) bands light curves for 
the 2006 outburst of recurrent nova RS Oph. 
%The plateau phase from day 40
%to day 75 is a contribution of the disk irradiated by the hot WD. A sharp final 
%decline from day 75 is a results of hydrogen burning turnoff. 
Theoretical model \cite{katomariko.hac06a} 
indicates that the WD of RS Oph is very massive, 1.35 $M_\odot$.  Figure taken from 
\cite{katomariko.hac06a}.}
    \label{katomariko.f2}   \end{center} \end{minipage}

% \begin{figure}[t]%%%%%%%%  PU Vul %%%%%%%% \begin{tabular}{cc}  
\begin{minipage}{.55\hsize}   \begin{center}
     \includegraphics[width=.98\textwidth]{f3.epsi}
      \caption{ Optical and UV light curves of PU Vul.
Large open circles denote the continuum flux of IUE UV 1455~\AA~ band.
The scale in the right-hand-side denotes that for observational
data. No optically thick wind occurs. 
%The cross/small dot indicate starting/end point of optically-thin wind
%mass-loss. 
Solid and dashed lines denote theoretical nova model of 0.57 and 0.6 
$M_\odot$ WD. The downward arrow indicates the epoch of eclipse. 
Taken from  \cite{katomariko.puvul10}.}
    \label{katomariko.f3}   \end{center}  \end{minipage} \end{tabular}
\end{figure}%%%%%%%%%%%%%%%

\section{Introduction} %%%%%%%%%%%%%%%

Nova is a thermonuclear runaway event on an accreting WD.
After a shell flash sets in the envelope of the WD expands to a giant size.
Strong optically thick winds occur that blow off a large part of the 
envelope. The photospheric temperature rises with time that 
causes the optical magnitude decrease while the total luminosity is 
almost constant.
Figure 1 shows the HR diagram for the decay phase of nova outburst of 
solar composition 
(see \cite{katomariko.kat94} for HR diagram of CO nova). 
The optically thick winds widely 
occur and continue until the small open circles in Figure 1 (in green region).

In the decay phase the envelope
settles down into a thermal equilibrium in which
nuclear energy generation is balanced with radiative loss. 
The star moves leftward as the envelope mass decreases due to 
wind mass-loss and nuclear burning. 
Finally hydrogen nuclear burning stops  and the star cools down. 

The decay timescale depends strongly on the WD mass. 
A very massive WD crosses HR diagram in a few month. 
For example, recurrent nova RS Oph ($M_{\rm WD} \sim 1.35 M_\odot$
\cite{katomariko.hac06b}) shows the total decay time 
as short as 80 days as shown in Figure 2. Much longer timescales are 
obtained for less massive WDs, e.g., 7 yr for $1.0 M_\odot$, 80 yr for 0.6 $M_\odot$, 270 yr 
for 0.5 $M_\odot$ \cite{katomariko.kat94}.
 
Figure 1 distinguishes the regions where the optically thick wind occurs or not. 
The wind does not occur in  the high 
temperature region of each WD and also in the less massive WDs.
As the optically thick wind is accelerated
deep inside the photosphere where the OPAL opacity has a prominent peak  
at $\log T $(K)$ \sim 5.2$ \cite{katomariko.kat94},  
no wind is accelerated when the photospheric 
temperature is higher than this peak temperature. 
This is the reason of no wind in the left side of the small dot in   
Figure 1. 
On the other hand, 
in less massive WDs ($\leq 0.5 M_\odot$), optically thick wind does not occur 
because the luminosity is always smaller than the 
local Eddington luminosity (described later)  
and the radiation-pressure-gradient is insufficient to accelerate the wind.

The occurrence of optically thick winds in low mass WDs ($\sim 0.6~M_\odot$) 
has not been fully clarified yet.
Kato and Hachisu (2009) examined the boundary condition of occurrence of the 
wind and found that there are two kind of solutions with/without winds, 
both of which can be realized in nova outbursts on low mass WDs 
as shown in Figure 1, 
and that the occurrence of the winds 
depends on the ignition mass of a shell flash. 
As the optically thick winds essentially determine a nova evolution 
because of its large mass-loss rate, it is important to know when
it occurs (when it does not) and how different the light curves are.

This short report aims to explain how and why the difference arises between 
the sharp peak such as in RS Oph (Figure 2) and other many classical nova 
and the long-lasted flat peak in PU Vul (Figure 3). 
This report is mainly based on the results of \cite{katomariko.occurrence}.
In Section 2, we introduce our simplified model to follow evolution of 
nova outbursts.
Internal structure and evolution of nova envelopes are 
explained in Section 3.
Section 4 presents the occurrence condition of optically thick winds 
for various WD masses that closely relate with the flat peak light curve.
 Conclusions follow.

\section{Simplified Model} \label{kato_sec_model}

We can approximate 
nova evolution with a sequence of solutions that consists of steady-wind and 
static solutions because evolution timescale is much larger than hydrodynamical 
timescale except very early phase of shell flash \cite{katomariko.kat89sh,
katomariko.kat94,katomariko.pri86}. 
When the optically thick wind occurs we solve the equations
of motion, mass continuity, radiative diffusion, and conservation of energy,
from the bottom of the hydrogen-rich envelope through the photosphere
assuming steady-state. 
When the optically thick wind does not occur, we solve 
equation of hydrostatic balance instead of equation of motion.
The occurrence of optically thick winds is detected by the conditions  
(1) the photospheric luminosity approaches the Eddington limit
and (2) at the same time the thermal energy at the photosphere is comparable
to the gravitational energy as described in Kato (1985) \cite{katomariko.kat85}.

In the rising phase of nova outburst, I integrated energy 
conservation equation without energy generation term due to nuclear 
burning and later estimated energy generation
using the temperature and density obtained. 
In the decay phase I set the condition that the energy generation is
balanced with radiative energy loss because the envelope settles down 
into a thermal equilibrium \cite{katomariko.pri86}.
Convective energy transport is calculated in static solutions using
the mixing length theory with $\alpha=1.5$. 
The OPAL opacity \cite{katomariko.opal} is used.
The chemical composition of the envelope is assumed to be uniform
throughout the envelope with the solar composition $X=0.7$ and $Z=0.02$. 
These equations and method of
calculations are already published in \cite{katomariko.kat94,katomariko.occurrence}.

\begin{figure}[t]%%%% (Fig 4 and 5) rising phase of wind and st  %%%%%%%%%
 \begin{tabular}{cc} 
% source paper/occurrence/fig_LrT/LrTM06dM423E-5.f5.wip
    \label{katomariko.f4}   
 \begin{minipage}{.45\hsize}   \begin{center}
     \includegraphics[width=.98\textwidth]{f4.epsi}
      \caption{Evolutionary change of the diffusive and Eddington luminosities
in the rising phase of a nova. The envelope mass is 
%$\Delta M_{\rm ig}=
$4.2 \times 10^{-5}M_\odot$. $l$ denotes the ratio of
the luminosity to the Eddington luminosity at the photosphere.
The optically thick winds occur in (d)-(f).
The small dot denotes the critical point 
\cite{katomariko.kat83a,katomariko.kat83b,katomariko.kat94}. 
% (Taken from \cite{katomariko.occurrence})
}
\end{center}  \end{minipage}
  \begin{minipage}{.45\hsize}   \begin{center}
     \includegraphics[width=.98\textwidth]{f5.epsi}
      \caption{Same as Figure 4, but for the envelope mass 
%$\Delta M_{\rm ig}=
$7.0 \times 10^{-5}~M_\odot$. Optically 
thick winds are not accelerated and all the solutions are static solution. 
(Taken from \cite{katomariko.occurrence})
Note that the super Eddington region at $\log T \sim 5.2$ is narrow in this 
diagram, but very wide in Figure 6.
}
    \label{katomariko.f5}
%%%% source paper/occurrence/fig_LrT/LrTM06dM7E-5.f7.wip
   \end{center}  \end{minipage} \end{tabular}

\end{figure}%%%%%%%%%%%%%%%

\begin{figure}[t]%%% Fig 6 and 7:  comparison of ST .vs. ML: strucM06compSTML   %%%%

 \begin{tabular}{cc}
  \begin{minipage}{.5\hsize}
   \begin{center}
     \includegraphics[width=.98\textwidth]{f6.epsi}
      \caption{Comparison of two solutions with the same photospheric
temperature of $\log T_{\rm ph}$ (K) $= 4.53$ in the decay phase of
a 0.6 $~M_{\odot}$ WD. {\it Thick line}: envelope solution
in the static sequence. The envelope mass is
$2.3 \times 10^{-5}~M_{\odot}$. {\it Thin line}: envelope solution
in the wind sequence. The envelope mass is $3.8 \times 10^{-5}~M_{\odot}$.
The right edge of each line corresponds to the photosphere.
(Taken from \cite{katomariko.occurrence})}
    \label{katomariko.f6}   \end{center}  \end{minipage}
%  source: aurora: paper/occurrence/fig/strucM06compSTML.f9.wip
% \end{tabular}
%\end{figure}%%%%%%%%%%%%%%%

%\begin{figure}[t]%%% Fig :  occurrence diagram of wind  %%%%%%%%%%%%

% \begin{tabular}{cc} 
 \begin{minipage}{.45\hsize}   \begin{center}
     \includegraphics[width=.98\textwidth]{f7.epsi}
      \caption{Upper and lower critical ignition masses for winds,
$\Delta M_{\rm exp}$ and  $\Delta M_{\rm wind}$, plotted against the
WD mass. The optically thick winds occur
in the right hand side of the lines (labeled ``wind'').
In the upper side of the line, optically thick
wind does not occur and the envelope expands quasi-statically
(labeled ``expansion'').
In the lower region, the envelope does not expand so much (``no expansion'') and
no wind arises. 
(Taken from \cite{katomariko.occurrence})
}
    \label{katomariko.f7}   \end{center}  \end{minipage}
%  source: paper/occurrence/fig_lsmaxdM/lsmaxdM.f10.wip
 \end{tabular}
\end{figure}%%%%%%%%%%%%%%%

\section{Internal Structures} \label{kato_sec_structure} %%%%%%%%%%%%%%%

Figures 4 and 5 show 
the distribution of the 
diffusive luminosity and the local Eddington luminosity against the
temperature for solutions along the rising phase.
Here, the local Eddington luminosity is defined as

\begin{equation}
L_{\rm Edd} \equiv {4\pi cGM \over\kappa},
\end{equation}

\noindent
where $\kappa$ is the OPAL opacity. 
Since the opacity $\kappa$ is a function of the temperature and density,
the Eddington luminosity is also a local variable.
This Eddington luminosity has a local minimum at $\log T$ (K) = 5.25
corresponding to the opacity peak as shown in e.g., Figures 4f and 5f.

As the shell flash goes on, the diffusive luminosity
increases with time and approaches the Eddington luminosity near the photosphere
(Figures 4 a,b, and c).
%  where, $l$ denotes the ratio of
% the luminosity to the Eddington luminosity at the photosphere.).
When the photospheric temperature decreases to $\log T$ (K) $\sim 5.2$,
matter is accelerated and optically thick steady wind begins (Figure 4d).
After that, the envelope continuously expands to reach the maximum expansion, 
where the optical magnitude reaches the maximum. 
A narrow super-Eddington region appears corresponding to the opacity peak
at $\log T$ (K) $\sim 5.2$.

Figure 5 shows a rising phase similar to those in Figure 4, but for the case of 
massive envelope of $\Delta M_{\rm ig}=7.0 
\times 10^{-5}~M_{\odot}$. In this case no winds are accelerated 
when the envelope expands beyond 
the opacity peak of $\log T$ (K) $\sim 5.2$ (Figure 5d) and 
the envelope continuously expands without optically thick winds (Figure 5e and 5f).  

Note that the structure of the static
solution just before the wind occurs (Figure 4c) is very similar to that
of the adjacent wind solution (Figure 4d) as pointed out by
\cite{katomariko.kat85} for the old opacity, 
which means that the optically thick winds 
occur very quietly. Moreover, the two solutions, Figure 4c 
and Figure 5c are also very similar. 
In later stages, however, the two sequences develop very different 
envelope structures.

Figure 6 compares internal structures of two types 
of solutions, with/without optically thick winds. These envelopes are for 
the same WD mass, chemical composition, and photospheric temperature.  
We see a remarkable difference in the density distribution.
The optically thick wind solution (thin line) shows 
monotonically decreasing density as  $r^{-2}$ in the outer envelope
($\log r$ (cm) $\geq 10.3$), while 
the static solution 
develops a large density-inversion region at $\log r$ (cm) $\sim 10-11.3$
corresponding to a super Eddington region.
This density-inversion arises in order to keep hydrostatic balance
in the super-Eddington region ($L_{\rm Edd} < L_r$) as expected
from the equation of hydrostatic balance.
Inefficient convections occur in the region of  $L_{\rm Edd} < L_r$ but are
unable to carry all of the diffusive energy flux, thus the structure is
super-adiabatic. 

In 0.4 and 0.5 $M_\odot$ WD, the envelope structure is more or less like 
that of the static solution of 0.6  $M_\odot$, i.e., with a wide density-inversion. 
However, in such less massive WDs, optically thick winds are not accelerated 
at all; No wind solutions exist.

\section{Occurrence of Winds and Sharp/Flat Optical Maximum}\label{kato_sec_occurrence} %%%%%%%

As we have seen in the previous sections, the optically thick wind occurs 
in a smaller envelope mass ($\Delta M_{\rm ig}=4.2 \times 10^{-5}~M_{\odot}$), 
but suppressed in a larger envelope mass ($7.0 
\times 10^{-5}~M_{\odot}$) for 0.6 $M_\odot$ WD.
In other words, for a given WD mass ($ > 0.5 M_\odot$), winds occur 
during a shell flash for a limited range of the ignition mass.

Figure 7 shows the region of the wind mass-loss in 
the diagram of the WD mass - ignition mass. 
When the ignition mass is smaller than $\Delta M_{\rm wind}$ the envelope
cannot expand much and the photospheric temperature
does not decrease to the opacity peak of $\log T$ (K) $\sim 5.2$
(see Figures 4a-c and 5a-c).
Therefore, no optically thick winds occur (``no expansion'').
On the other hand, when the ignition mass is larger than $\Delta M_{\rm exp}$,
winds are suppressed in a way that density-inversion
balances to radiation-pressure gradient in a super-adiabatic region 
(the region ``expansion'' which is an abbreviation of ``quasi-static expansion'').
In this case the envelope expands quietly because of no winds.
Therefore, optically thick winds occur only for
$\Delta M_{\rm wind} < \Delta M_{\rm ig} < \Delta M_{\rm exp}$.

The upper part of Figure 7, however, is not realized in actual mass-accreting WDs. 
Numerical calculations of shell flash showed that the ignition 
mass is up to 
$6\times 10^{-7}~M_{\odot}$ for 1.4 ~$M_{\odot}$, 
$8\times 10^{-5}~ M_{\odot}$ for a 1.0 ~$M_{\odot}$ WD, and 
$3\times 10^{-4}~ M_{\odot}$ for a 0.65 ~$M_{\odot}$ WD \cite{katomariko.pri95} 
for various accretion rates.
Therefore, ``expansion''  solutions are realistic only 
in less massive WDs, $\sim 0.5-0.7~M_{\odot}$.
When the ignition mass is larger than $\Delta M_{\rm exp}$ the shell flash 
develops  in a quasi-static manner. When the ignition mass is smaller, 
the shell flash evolves as a usual nova outburst. A small difference in  
ignition mass causes large difference in the envelope structure as well as
the evolution timescale, because the strong wind mass-loss governs nova 
evolution once it occurs.  

Kato et al. (2010) \cite{katomariko.puvul10} show that such 
a ``static expansion'' is realized 
in symbiotic nova, PU Vul, which showed no indication of strong wind mass-loss 
in its spectrum during the flat peak. 
Figure 3 shows the theoretical 
models calculated by Kato et al. for $\sim 0.6 M_\odot$ WDs. 
As no optically thick winds occur, the evolution timescale 
is as long as a decade, thus the envelope keeps low surface temperatures  
several years. This is the reason why PU Vul shows a long-lasted flat 
optical maximum. This is a remarkable contrast with a majority of novae 
that shows a sharp peak like RS Oph (Figure 2), 
in which strong optically thick winds blow off a large part of the envelope 
in a very short timescale.

\section{Conclusion} %%%%%%%%%%%%%%%

Once the optically thick wind occurs, it blows off a large part of the 
envelope mass, thus the light curve quickly decreases immediately 
after the optical maximum, making a sharp optical peak. 
When the winds does not occur, the light curve shows a  
long-lasted flat peak. Such a flat peak will be observed only in less massive WDs 
($< 0.7 M_\odot$), 
because wind-suppressed solutions are realized only in less massive WDs  
with a relatively massive ignition mass. PU Vul is a good example of such 
a wind-suppressed evolution.

%%%%%%%%%%%%%%%%%%%%%%%%%%%%%%%%%%
%% thebibliography environment %%
%%%%%%%%%%%%%%%%%%%%%%%%%%%%%%%%%
% todayref

%%%%%%%%%%
\end{document}